\documentclass[twocolumn,twocolappendix]{aastex63}

\received{}
\revised{}
\accepted{}

\submitjournal{The Astrophysical Journal}

\shorttitle{Electron Firehose Instabilities in High-$\beta$ ICM shocks}
\shortauthors{Kim et al.}

\begin{document}

\title{Electron Firehose Instabilities in High-$\beta$ Intracluster Shocks}

\author{Sunjung Kim}
\affil{Department of Physics, School of Natural Sciences UNIST, Ulsan 44919, Korea}
\author[0000-0001-7670-4897]{Ji-Hoon Ha}
\affil{Department of Physics, School of Natural Sciences UNIST, Ulsan 44919, Korea}
\author[0000-0002-5455-2957]{Dongsu Ryu}
\affil{Department of Physics, School of Natural Sciences UNIST, Ulsan 44919, Korea}
\author[0000-0002-4674-5687]{Hyesung Kang}
\affil{Department of Earth Sciences, Pusan National University, Busan 46241, Korea}

\correspondingauthor{Dongsu Ryu}
\email{ryu@sirius.unist.ac.kr}

\begin{abstract}

The preacceleration of electrons through reflection and shock drift acceleration (SDA) is essential for the diffusive shock acceleration (DSA) of nonthermal electrons in collisionless shocks. Previous studies suggested that, in weak quasi-perpendicular ($Q_\perp$) shocks in the high-$\beta$ ($\beta=P_{\rm gas}/P_{\rm B}$) intracluster medium (ICM), the temperature anisotropy due to SDA-reflected electrons can drive the electron firehose instability, which excites oblique nonpropagating waves in the shock foot. In this paper, we investigate, through a linear analysis and particle-in-cell (PIC) simulations, the firehose instabilities driven by an electron temperature anisotropy (ETAFI) and also by a drifting electron beam (EBFI) in $\beta\sim100$ ICM plasmas. The EBFI should be more relevant in describing the self-excitation of upstream waves in $Q_\perp$-shocks, since backstreaming electrons in the shock foot behave more like an electron beam rather than an anisotropic bi-Maxwellian population. We find that the basic properties of the two instabilities, such as the growth rate, $\gamma$, and the wavenumber of fast-growing oblique modes are similar in the ICM environment, with one exception; while the waves excited by the ETAFI are nonpropagating ($\omega_r=0$), those excited by the EBFI have a non-zero frequency ($\omega_r\neq0$). However, the frequency is small with $\omega_r<\gamma$. Thus, we conclude that the interpretation of previous studies for the nature of upstream waves based on the ETAFI remains valid in $Q_\perp$-shocks in the ICM.

\end{abstract}


\section{Introduction}\label{sec:s1}

In the current $\Lambda$CDM cosmology, galaxy clusters emerge through hierarchical clustering, and the ensuing supersonic flow motions generate shock waves in the intracluster medium (ICM) \citep[e.g.,][]{miniati2000,ryu2003,pfrommer2006,skillman2008,vazza2009,hong14,schaal2015}. Among the ICM shocks, the merger shocks, induced during mergers of sub-clusters, are the most energetic; they form in almost virial equilibrium and hence are weak with low sonic Mach numbers of $M_{\rm s}\lesssim4$ \citep[e.g.,][]{ha2018}. As in most astrophysical shocks, cosmic ray (CR) protons and electrons are expected to be accelerated via diffusive shock acceleration (DSA) in these ICM shocks \citep[e.g.,][]{bell1978,blandford1978, drury1983,brunetti2014}. From observations of the so-called radio relics \citep[e.g.,][]{vanweeren2010,vanweeren2012}, in particular, the electron acceleration is inferred to operate in low Mach number, quasi-perpendicular ($Q_\perp$, hereafter) shocks with $\theta_{\rm Bn}\gtrsim45^{\circ}$ in the hot ICM \citep[see][for a review]{vanweeren2019}. Here, $\theta_{\rm Bn}$ is the obliquity angle between the background magnetic field and the shock normal.

The preacceleration of electrons is one of the long-standing unsolved problems in the theory of DSA \citep[e.g.,][]{marcowith2016}. Thermal electrons need to be energized to suprathermal energies ($p_e\gtrsim{\rm a~few}\times p_{\rm th,p}$) in order to be injected to the DSA process, because the full DSA requires the diffusion of CR electrons both upstream and downstream across the shock and the width of the shock transition region is comparable to the ion gyroradius. Here, $p_{\rm th,p}=(2m_p k_B T_{2})^{1/2}$ is the proton thermal momentum in the postshock gas of temperature $T_{2}$, $m_p$ is the proton mass, and $k_B$ is the Boltzmann constant. The electron injection, which is known to be effective mainly at $Q_\perp$-shocks \citep[e.g.,][]{gosling1989,burgess2007}, involves the following key elements:
(1) the reflection of incoming ions and electrons at the shock ramp due to magnetic mirror forces, leading to backstreaming,
(2) the energy gain from the motional electric field in the upstream region through shock drift acceleration (SDA) or shock surfing acceleration (SSA), and/or through interactions with waves,
and (3) the trapping of electrons near the shock due to the scattering by the upstream waves, excited by backstreaming ions and electrons, which allows multiple cycles of SDA or SSA \citep{amano2009,riquelme2011,matsukiyo11,matsumoto2012,guo2014a,guo2014b,kang2019}.
The injection problem can be followed from first principles only through particle-in-cell (PIC) simulations, which fully treat kinetic microinstabilities and wave-particle interactions on both ion and electron scales around the shock transition.

Self-generated upstream waves play an important role in the electron preacceleration, because in the absence of scattering by the upstream waves, the energization of electrons would be terminated after one SDA cycle. Plethora of plasma instabilities can be destabilized in the shock foot, depending on the shock parameters, such as the plasma beta, $\beta=P_{\rm gas}/P_{\rm B}$ (the ratio of the gas to magnetic pressures), the sonic Mach number $M_{\rm s}$, the Alfv\'en Mach numbers, $M_{\rm A}$ ($M_{\rm A}=\sqrt{\beta\Gamma/2}M_{\rm s}$ where $\Gamma=5/3$ is the gas adiabatic index), the obliquity angle, $\theta_{\rm Bn}$, and the adopted ion-to-electron mass ration, $m_p/m_e$\footnote{We use the term `ion' to represent the positively charged particle with a range of $m_p/m_e=100-1836$. But we use the subscript $p$ to denote the ion population, since the subscript $i$ is used for the coordinate component.} \citep[e.g.][]{matsukiyo06,balogh2013}. Early PIC simulation studies centered on high Mach number shocks in $\beta\sim 1$ plasmas, characterizing the Earth bow shock in the solar wind and supernova blast waves in the interstellar medium. For instance, \citet{amano2009} and \citet{matsumoto2012} found that at high $M_{\rm A}$ (specifically, $M_{\rm A}>(m_p/m_e)^{2/3}$) $Q_\perp$-shocks in $\beta\approx1$ plasmas, the drift between reflected ions and incoming electrons triggers the Buneman instability, which excites large-amplitude electrostatic waves. Then, backstreaming electrons are scattered by these waves and gain energies via multiple cycles of SSA at the leading edge of the shock foot. On the other hand, \citet{riquelme2011} argued that at low $M_{\rm A}$ (with $M_{\rm A}\lesssim(m_p/m_e)^{1/2}$) $Q_\perp$-shocks in $\beta\approx1$ plasmas, modified two-stream instabilities can excite oblique whistler waves and electrons gain energies via wave-particle interactions with those whistlers in the shock foot. \citet{matsukiyo11} also considered $M_{\rm A}\approx 5-8$, $Q_\perp$-shocks in $\beta\approx 3$ plasmas, and showed that reflected electrons are energized through SDA.

The ICM, which is composed of the hot gas of a few to several keV and the magnetic fields of the order of $\mu$G, on the other hand, contains plasmas of $\beta\sim100$ \citep[e.g.,][]{ryu2008,brunetti2014,porter2015}. Hence, although ICM shocks are weak with low sonic Mach numbers of $M_{\rm s} \lesssim 4$, they have relatively high Alfv\'en Mach numbers of up to $M_{\rm A} \lesssim 40$. To understand the electron acceleration at ICM shocks, \citet{guo2014a,guo2014b} performed two-dimensional (2D) PIC simulations of $M_{\rm s}=3$, $Q_\perp$-shocks in $\beta= 6-200$ plasmas with $k_B T=86$ keV. They argued that the temperature anisotropy ($T_{e\parallel}>T_{e\perp}$) due to reflected electrons, backstreaming along the background magnetic fields with small pitch angles, derives the electron firehose instability (EFI, hereafter), which excites mainly nonpropagating oblique waves in the shock foot. Here, $T_{e\parallel}$ and $T_{e\perp}$ are the electron temperatures, parallel and perpendicular to the background magnetic field, respectively. The SDA-reflected electrons\footnote{The SDA-reflected electrons mean those that are energized by SDA in the course of reflection.} are scattered back and forth between the magnetic mirror at the shock ramp and the EFI-driven upstream waves, but they are still suprathermal and do not have sufficient energies to diffuse downstream across the shock transition. On the other hand, \citet{trotta2019} and \citet{kobzar2019} have recently shown through 2D and 3D plasma simulations of supercritcal $Q_\perp$-shocks that shock surface ripplings generate multi-scale perturbations that can facilitate the electron acceleration beyond the injection momentum.

\citet[KRH19, hereafter]{kang2019} revisited this problem with wider ranges of shock parameters that are more relevant to ICM shocks, $M_{\rm s}=2.0-3.0$, $\beta=50-100$, and $k_B T=8.6$ keV. They showed that the electron preacceleration through the combination of reflection, SDA, and EFI may operate only in {\it supercritical}, $Q_\perp$-shocks with $M_{\rm s} \gtrsim 2.3$.
In addition, they argued that the EFI alone may not energize the electrons all the way to the injection momentum, $p_{\rm inj} \sim 130 p_{\rm th,e}$ (where $p_{\rm th,e}=(2m_e k_B T_{2})^{1/2}$), unless there are pre-existing turbulent waves with wavelengths longer than those of the EFI-driven waves ($\lambda \gtrsim 20 c/ \omega_{pe}$, where $c$ is the speed of light and $\omega_{pe}$ is the electron plasma frequency). Analyzing self-excited waves in the shock foot, they found that (1) nonpropagating oblique waves with $\lambda \sim 15-20 c/ \omega_{pe}$ are dominantly excited, (2) the waves decay to those with longer wavelengths and smaller propagation angle, $\theta=\cos^{-1}(\mathbf{k}\cdot\mathbf{B}_0/kB_0)$, the angle between the wavevector, $\mathbf{k}$, and the background magnetic field $\mathbf{B}_0$, and (3) the scattering of electrons by those waves reduces the temperature anisotropy. These findings are consistent with previous works on EFI using linear analyses and PIC simulations \citep[e.g.,][]{gary2003,camporeale2008,hellinger2014}.

The EFI in homogeneous, magnetized, collisionless plasmas has been extensively studied in the space-physics community as a key mechanism that constrains the electron anisotropy in the solar wind \citep[see][for a review]{gary1993}. It comes in the following two varieties: (1) the electron temperature-anisotropy firehose instability (ETAFI, hereafter), driven by a temperature anisotropy, $T_{e\parallel} > T_{e\perp}$ \citep[e.g.][]{gary2003,camporeale2008,hellinger2014}, and (2) the electron beam firehose instability (EBFI, hereafter), also known as the electron heat flux instability, induced by a drifting beam of electrons \citep[e.g.][]{gary1985,saeed2017,shaaban2018}. In the EBFI, the bulk kinetic energy of electrons is the free energy that drives the instability. In the linear analyses of these instabilities, typically ions are represented by an isotropic Maxwellian velocity distribution function (VDF, hereafter) with $T_{\rm p}$, while electrons have different distributions, that is, either a single anisotropic bi-Maxwellian VDF with $T_{e\parallel}>T_{e\perp}$ for the ETAFI, or two isotropic Maxwellian VDFs (i.e., the core with $T_{\rm c}$ and the beam with $T_{\rm b}$) with a relative drift speed, $u_{\rm rel}$, for the EBFI.

The main findings of the previous studies of the ETAFI can be summarized as follows \citep[see, e.g.,][]{gary2003}. (1) The threshold condition of the instability decreases with increasing $\beta_{\rm e}$, approximately as $(T_{e\parallel}-T_{e\perp})/T_{e\parallel}\gtrsim 1.3~\beta_{\rm e}^{-1}$. (2) The instability induces two branches, i.e., the parallel ($\theta\approx0^{\circ}$), nonresonant, propagating ($\omega_r \ne 0$) mode and the oblique ($\theta\gg0^{\circ}$), resonant, nonpropagating ($\omega_r=0$) mode. The propagating mode is left-hand (LH) polarized. The latter, nonpropagating mode has the growth rate higher than the former. (3) The perturbed magnetic field of the nonpropagating mode is dominantly along the direction perpendicular to both $\mathbf{k}$ and $\mathbf{B_0}$. (4) The oblique nonpropagating modes decay to the propagating modes of smaller wavenumbers and smaller angles. (5) The ETAFI-induced waves scatter electrons, resulting in the reduction of the electron temperature anisotropy and the damping of the waves.

For the case of the electron heat flux instability (i.e., the EBFI), the parallel-propagating ($\theta=0^{\circ}$) mode was analyzed before, and is known to have two branches, that is, the right-hand (RH) polarized whistler mode and the LH polarized firehose mode \citep{gary1993}. The firehose mode becomes dominant at sufficiently large drift speeds \citep{gary1985}. Although the oblique ($\theta\ne0^{\circ}$) mode has not been sufficiently examined in the literature so far, it is natural to expect that the oblique mode would have the growth rate larger than the parallel mode, similarly as in the case of the ETAFI \citep{saeed2017}. Recently, \citet{shaaban2018} studied the electron heat flux instability driven by a drifting beam of anisotropic bi-Maxwellian electrons, but again only for the parallel propagation.

As mentioned above, \citet{guo2014b} and KRH19 argued that the upstream waves in the shock foot in their PIC simulations have the characteristics consistent with the nonprogating oblique waves excited by the ETAFI. Considering that backstreaming electrons would behave like a drifting beam, however, it would have been more appropriate to interpret the operating instability as the EBFI. So we here consider and compare the two instabilities, in order to understand the nature of the upstream waves in $Q_\perp$-shocks in high-$\beta$ plasmas. Another reason why we study this problem is that the ETAFI and EBFI in high-$\beta$ plasmas have not been examined before. In particular, we study the instabilities at both parallel and oblique propagations through the kinetic Vlasov linear theory and 2D PIC simulations, focusing on the kinetic properties of the EFI in high-$\beta$ ($\beta_{\rm p}\approx 50$ and $\beta_{\rm e}\approx 50$) plasmas relevant for the ICM.

The paper is organized as follows. Section \ref{sec:s2} describes the linear analysis of the ETAFI and EBFI. In Section \ref{sec:s3}, we present the nonlinear evolution of the EBFI in 2D PIC simulations in a periodic box. A brief summary is given in Section \ref{sec:s4}. 

\section{Linear Analysis of ETAFI and EBFI}\label{sec:s2}

We consider the ETAFI and EBFI in a homogeneous, collisionless, magnetized plasma, which is specified by the density and temperature of ions and electrons, $n_p$, $n_e$, $T_p$, $T_e$, and the background magnetic field of $\mathbf{B}_0$. For the case of the ETAFI, the anisotropic bi-Maxwellian distribution of electrons is described by $T_{e\parallel}$ and $T_{e\perp}$, and the temperature anisotropy parameter is given as $\mathcal{A}=T_{e\parallel}/T_{e\perp}$. The ion population is described with a single temperature. For the case of the EBFI, the core and beam populations of electrons with the drift speeds of $u_c$ and $u_b$ along the direction of the background magnetic field are assumed\footnote{The subscripts $c$ and $b$ stand for the core and beam populations.}. The ion population is on average at rest with zero drift speed. The adopted ion-to-electron mass ratio includes the realistic ratio, $m_p/m_e=1836$, of the proton to electron mass and a reduced one, $m_p/m_e=100$. The reduced mass ratio is considered for comparison with the PIC simulations in the next section where $m_p/m_e=100$ and 400 are adopted.

The VDF of a drifting bi-Maxwellian population can be written in the general form,
\begin{equation}
\label{eq:e01}
f_a(v_{\perp},v_{\parallel}) = \frac{n_a}{n_0}\frac{\pi^{-3/2}}{\alpha_{a \perp}^2\alpha_{a \parallel}} \exp\left[-\frac{v_{\perp}^2}{\alpha_{a \perp}^2}-\frac{(v_{\parallel}-u_a)^2}{\alpha_{a \parallel}^2}\right].
\end{equation}
The subscript $a$ can denote the population of core electrons ($c$), beam electrons ($b$), or ions ($p$). Here, $n_a$ is the number density of the particle species $a$, and $n_0$ is the number density of electrons and ions, which satisfy $n_0=n_c+n_b=n_p$, the charge neutarlity condition; $u_a$ is the drift speed directed along the background magnetic field, and satisfies $n_c u_c + n_b u_b - n_p u_p=0$, the zero net current condition. The thermal velocities are $\alpha_{a \parallel}=\sqrt{2 k_{\rm B} T_{a \parallel}/m_a}$ and $\alpha_{a \perp}=\sqrt{2 k_B T_{a \perp}/m_a}$, respectively. Throughout the paper, the plasma beta, $\beta_a=8\pi n_0 k_B T_{a}/B_0^2$, the plasma frequency, $\omega_{pa}^2=4\pi n_0 e^2/m_a$, and the gyro-frequency, $\Omega_a=e B_0/m_a c$, for electrons and ions are used. The Alfv$\rm{\acute{e}}$n speed, given as $v_A=(B_0^2/4\pi n_0 m_p)^{1/2}$, is also used. Note that for the ion (proton) population,  $T_{p\parallel}=T_{p\perp}=T_p$ in the ETAFI analysis, while $u_p=0$
in the EBFI analysis in the following subsections.

The linear dispersion relation of general electromagnetic (EM) modes for the ETAFI and EBFI can be derived from the normal mode analysis with the linearized Vlasov-Maxwell system of equations for plasmas of multi-species. The derivation can be found in standard textbooks on plasma physics \citep[e.g.,][]{stix1992,brambilla1998}. The dispersion relation is given as
\begin{equation}
\label{eq:e02}
\det \left(\epsilon_{ij}-\frac{c^2 k^2}{\omega^2}\big(\delta_{ij}-\frac{k_i k_j}{k^2}\big) \right)=0,
\end{equation}
with the dielectric tensor, $\epsilon_{ij}$, where $k_i$ and $k_j$ are the components of the wavevector $\mathbf{k}$. Then, the complex frequency, $\omega = \omega_r + i \gamma$,\footnote{The quantity $i$ is the imaginary unit, not the coordinate component.} can be calculated as a function of the wave number, $k$, and the propagation angle, $\theta$. The dielectric tensor for the general VDF is given in the appendix. Setting that $\parallel$ is the $z$-direction and $\perp$ is the $x$-direction without loss of generality, that is, $\mathbf{B}_0=B_0{\hat z}$ and $\mathbf{k}$ are in the $z-x$ plane, the components of $\epsilon_{ij}$ for the VDF in Equation (\ref{eq:e01}) is written as
\begin{eqnarray}
\label{eq:e03}
\epsilon_{xx} &=& 1+\sum_{a=c,b,p}\frac{\omega_{pa}^2}{\omega^2}\frac{n_a}{n_0}\sum_{n=-\infty}^{n=\infty}
\frac{n^2 \Lambda_n (\lambda_a)}{\lambda_a}A_n^a,
\nonumber \\
\epsilon_{yy} &=& 1+\sum_{a=c,b,p}\frac{\omega_{pa}^2}{\omega^2}\frac{n_a}{n_0}\sum_{n=-\infty}^{n=\infty}
\left(\frac{n^2 \Lambda_n (\lambda_a)}{\lambda_a}-2\lambda_a \Lambda_n^\prime(\lambda_a)\right)A_n^a,
\nonumber \\
\epsilon_{zz} &=& 1-\sum_{a=c,b,p}\frac{\omega_{pa}^2}{\omega^2}\frac{n_a}{n_0}\frac{T_{a \parallel}}{T_{a \perp}}
\nonumber \\
&& \times \sum_{n=-\infty}^{n=\infty}\Lambda_n (\lambda_a)\left(\zeta_n^a B_n^a+2\frac{u_a}{\alpha_{a \parallel}}B_n^a-2\frac{u_a^2}{\alpha_{a \parallel}^2}A_n^a\right),
\nonumber \\
\epsilon_{xy} &=& -\epsilon_{yx} = i \sum_{a=c,b,p}\frac{\omega_{pa}^2}{\omega^2}\frac{n_a}{n_0}\sum_{n=-\infty}^{n=\infty} n\Lambda_n^{\prime}(\lambda_a) A_n^a,
\nonumber \\
\epsilon_{xz} &=& \epsilon_{zx} = -\sum_{a=c,b,p}\frac{\omega_{pa}^2}{\omega^2}\frac{n_a}{n_0}
\frac{k_{\perp} \alpha_{a \parallel}}{2\Omega_a} 
\nonumber \\
&& \quad\quad\quad \times \sum_{n=-\infty}^{n=\infty}
\frac{n \Lambda_n (\lambda_a)}{\lambda_a} \left(B_n^a-2\frac{u_a}{\alpha_{a \parallel}}A_n^a\right),
\nonumber \\
\epsilon_{yz} &=& -\epsilon_{zy} = i\sum_{a=c,b,p}\frac{\omega_{pa}^2}{\omega^2}\frac{n_a}{n_0}
\frac{k_{\perp} \alpha_{a \parallel}}{2\Omega_a} 
\nonumber \\
&& \quad\quad\quad \times \sum_{n=-\infty}^{n=\infty}
\Lambda_n^{\prime}(\lambda_a) \left(B_n^a-2\frac{u_a}{\alpha_{a \parallel}}A_n^a\right),
\end{eqnarray}
where
\begin{eqnarray}
\label{eq:e04}
&& \Lambda_n (\lambda_a)=I_n (\lambda_a) e^{-\lambda_a},
\nonumber \\
&& \lambda_a=\frac{k_{\perp}^2 \alpha_{a \perp}^2}{2 \Omega_a^2},
\nonumber \\
&& A_n^a=\zeta_0^a Z(\zeta_n^a)-\left(\frac{T_{a \perp}}{T_{a \parallel}}-1\right) \frac{Z^{\prime}(\zeta_n^a)}{2},
\nonumber \\
&& B_n^a=\left[\zeta_0^a-\left(\frac{T_{a \perp}}{T_{a \parallel}}-1\right)\zeta_n^a \right]Z^{\prime}(\zeta_n^a),
\nonumber \\
&& \zeta_n^a=\frac{\omega-n\Omega_a-k_{\parallel} u_a}{k_{\parallel} \alpha_{a \parallel}}.
\end{eqnarray}
Here, $I_n (\lambda)$ denotes the modified Bessel function of the first kind and $Z(\zeta)$ is the plasma dispersion function (see the appendix).
The prime in $\Lambda_n^\prime(\lambda)$ and $Z^{\prime}(\zeta)$ indicates the derivative with respect to the argument.

\begin{figure*}[t]
\vskip 0.2cm
\hskip 0cm
\centerline{\includegraphics[width=1\textwidth]{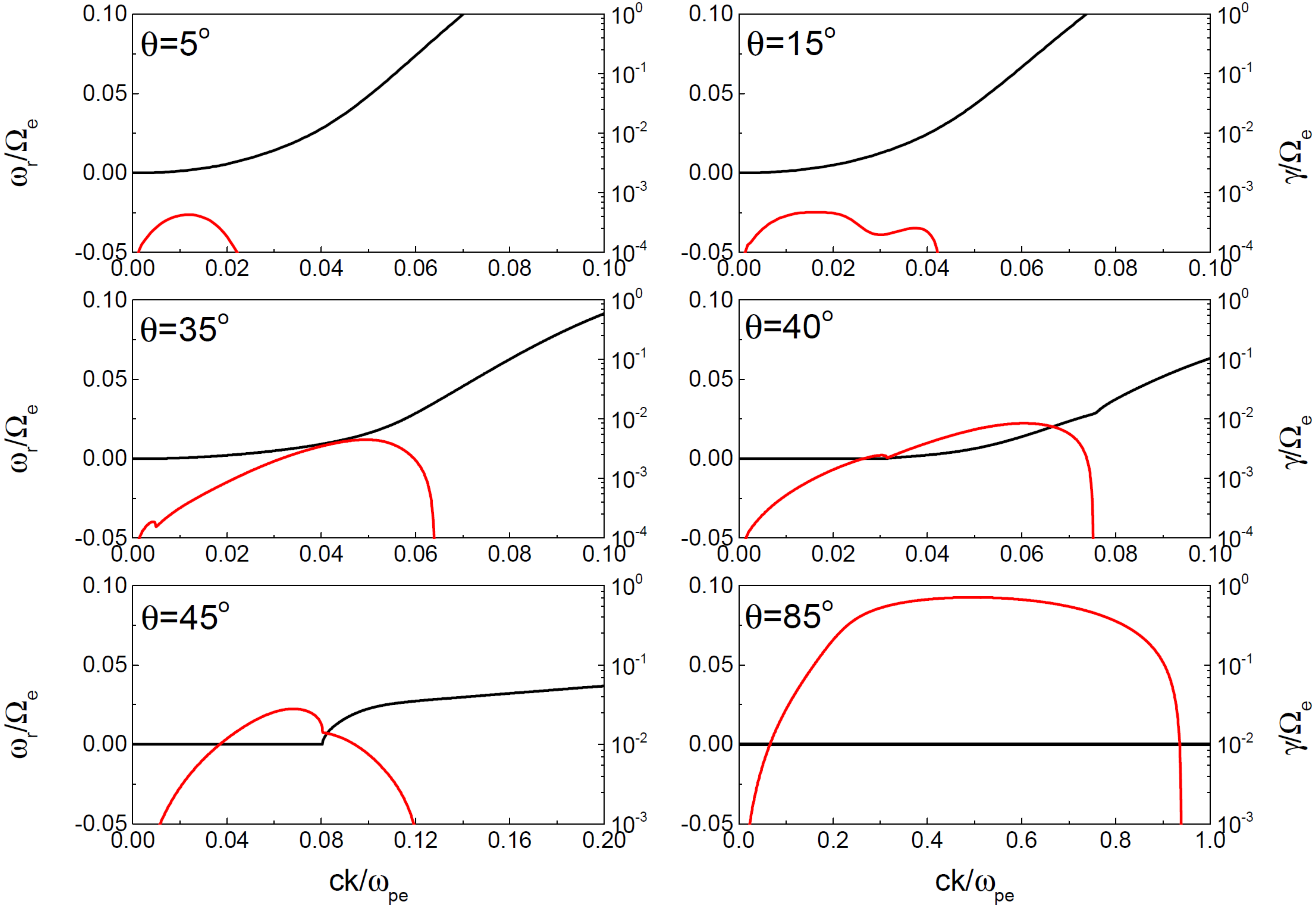}}
\vskip -0.2cm
\caption{Real frequency, $\omega_r$ (black), and growth rate, $\gamma$ (red), of the ETAFI as a function of wavenumber, $k$, for different propagation angle $\theta$, the angle between the wavevector and the background magnetic field. Here, $\mathcal{A}=T_{e\parallel}/T_{e\perp}=1.86$, $\beta=100$ (i.e., $\beta_{e \parallel}=72.3$, $\beta_{e \perp}=38.9$, and $\beta_p=50$), $v_A/c=6\times10^{-4}$, and $m_p/m_e=1836$. Note that the ranges of abscissa and ordinate differ in different panels.
\label{fig:f1}}
\end{figure*}

\begin{figure*}[t]
\vskip 0.2cm
\hskip 0cm
\centerline{\includegraphics[width=1\textwidth]{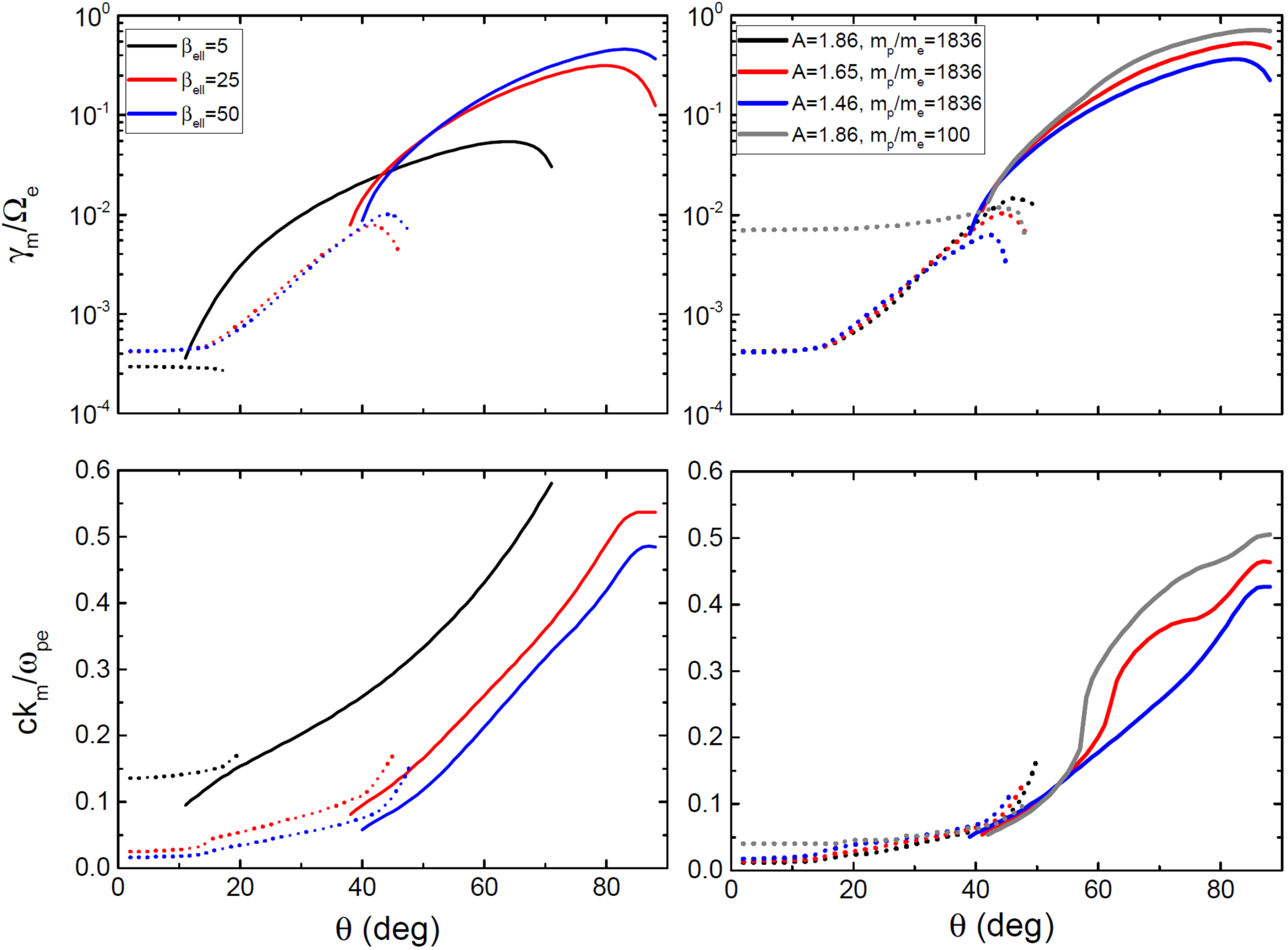}}
\vskip -0.2cm
\caption{Maximum growth rate, $\gamma_m$ (top), and wavenumber, $k_m$ (bottom), for the propagating (dashed line) and nonpropagating (solid line) modes of the ETAFI, as a function of $\theta$. The left panels are for different $\beta_{e\parallel}$ with a fixed $\mathcal{A}=1.67$; $\beta=2\beta_{e\parallel}(1+2/\mathcal{A})/3$, $v_A/c=10^{-4}$, and $m_p/m_e=1836$. The right panels are for different $\mathcal{A}$ and $m_p/m_e$; $\beta=100$ and $v_A/c=6\times10^{-4}/[(m_p/m_e)/1836]^{1/2}$ In the right panels, the gray solid lines almost completely overlap with the black solid lines.
\label{fig:f2}}
\end{figure*}

\subsection{Electron Temperature Anisotropy Firehose Instability (ETAFI)}\label{sec:s2.1}

We first examine the ETAFI in the ICM environment, triggered by the temperature anisotropy of electrons; hence, $T_{e\parallel}> T_{e\perp}$, while $T_{p\parallel}=T_{p\perp}$, and $u_c=u_b=0$ (no drift of electrons). For the anisotropic distribution of electrons in Equation (\ref{eq:e01}), the plasma beta is given as $\beta_{e\perp}=\beta_{e\parallel}/\mathcal{A}$ and $\beta_e= (\beta_{e\parallel} +2 \beta_{e\perp})/3 =\beta_{e\parallel} (1 + 2/\mathcal{A})/3$. We restrict the analysis to the case of $\beta_e=\beta_p$\footnote{The analysis can be easily extended to the general case of $\beta_e\ne\beta_p$.}. Then, the analysis is reduced to a problem of five parameters, for instance, $T_{e\parallel}$, $T_{e\perp}$, $n_0$, $B_0$ and $m_p/m_e$. We specify the problem with four dimensionless quantities, $\mathcal{A}=T_{e\parallel}/T_{e\perp}$, $\beta=\beta_e+\beta_p$, $v_A/c$, and $m_p/m_e$, and use $\omega_{pe}$ to normalize $k$. We then calculate $\omega_r/\Omega_e + i \gamma/\Omega_e$ as a function of $ck/\omega_{pe}$ and $\theta$. Note that $\Omega_e$ is given as a combination of other quantities, $\Omega_e=\omega_{pe}(v_A/c)(m_p/m_e)^{1/2}$. Considering that $n_0 \sim 10^{-4}\ {\rm cm^{-3}}$, $T_e \sim T_p\sim 10^8$~K (8.6 keV), and $B_0$ is of the order of $\mu{\rm G}$ in the ICM (see the introduction), we adopt $\beta=100$ and $v_A/c = [(2/\beta_p)(k_BT/m_pc^2)]^{1/2} \equiv 6\times10^{-4}/[(m_p/m_e)/1836]^{1/2}$ as fiducial values. 

Figure \ref{fig:f1} shows the analysis results of the ETAFI for the model of $\mathcal{A}=1.86$, $\beta=100$, $v_A/c=6\times10^{-4}$, and $m_p/m_e=1836$. The normalized real frequency, $\omega_r/\Omega_e$ (black line), and the normalized growth rate, $\gamma/\Omega_e$ (red line), are plotted as a function of $ck/\omega_{pe}$ for different $\theta$. At small, quasi-parallel angles ($\theta<35^{\circ}$), the propagating mode with $\omega_r\ne0$ dominates over the nonpropagating mode with $\omega_r=0$ in all the range of $k$. As $\theta$ increases, $\gamma$ of both the modes increase, but $\gamma$ of the nonpropagating mode increases more rapidly than that of the propagating mode (see the panels of $\theta=35^{\circ}$, $40^{\circ}$, and $45^{\circ}$). As $\theta$ increases further, the nonpropagating mode dominates in all the range of $k$ (see the panel of $85^{\circ}$). The maximum growth rate, $\gamma_m$, appears at $ck_m/\omega_{pe}\approx0.49$ and $\theta_m \approx85^{\circ}$, while nonpropagating modes with a broad range of $ck/\omega_{pe}\sim0.2-0.8$ have similar $\gamma$. This agrees with the previous finding that the ETAFI predominantly generates oblique phase-standing waves (see the introduction).

Figure \ref{fig:f2} shows the analysis results for different parameters. The left panels exhibit the dependence on $\beta_{e\parallel}$ with $\beta_{e\parallel}=5$, 25, and 50 for a fixed $\mathcal{A}=1.67$; then, $\beta = 7.33$, 36.6, and 73.3, respectively, and other parameters are $v_A/c=10^{-4}$ and $m_p/m_e=1836$. The maximum growth rate, $\gamma_m$, and the corresponding wavenumber, $k_m$, for given $\theta$, are plotted as a function of $\theta$ for both the propagating (dashed line) and nonpropagating (solid line) modes. The results of the $\beta_{e \parallel}=5$ model are in perfect agreement with the solutions provided by \cite{gary2003} (see their Figure 2), demonstrating the reliability of our analysis. The peak of $\gamma_m$ occrus at the nonpropagating mode, again indicating that the fastest-growing mode is nonpropagating, regardless of $\beta$. For higher $\beta$, the peak is higher\footnote{For higher $\beta$, $\Omega_e$, the normalization factor in the plot, is smaller, if the difference in $\beta$ is due to the difference in $B_0$,} and appears at larger $\theta$ and smaller $k$; that is, for higher $\beta$, the ETAFI grows faster, and the fastest-growing mode has a longer wavelength and a larger propagation angle.

The right panels of Figure \ref{fig:f2} examine the dependence on $\mathcal{A}$ in the range of $\mathcal{A}=1.46-1.86$ for a fixed $\beta=100$; then, $\beta_{e\parallel}=63.3-72.3$, and other parameters are $v_A/c=6\times10^{-4}$ and $m_p/m_e=1836$. For larger $\mathcal{A}$, the peak of $\gamma_m$ is higher and appears at larger $\theta$ and larger $k$; that is, for larger $\mathcal{A}$, the ETAFI grows faster, and the fastest-growing mode has a shorter wavelength and a larger propagation angle. The right panels also compare the models of $m_p/m_e=1836$ and 100 for $\mathcal{A}=1.86$. While the growth rate of the propagating mode strongly depends on $m_p/m_e$, the characteristics of the nonpropagating mode is insensitive to $m_p/m_e$ once $m_p/m_e$ is sufficiently large. This is consistent with the findings of \citet{gary2003}.

\begin{deluxetable*}{ccccccccccc}[t]
\tablecaption{Model Parameters for the Linear Analysis of the EBFI\label{tab:t1}}
\tabletypesize{\small}
\tablecolumns{10}
\tablenum{1}
\tablewidth{0pt}
\tablehead{
\colhead{Model Name $^{\rm a}$} &
\colhead{$\beta_e=\beta_p$} &
\colhead{$n_b/n_0$} &
\colhead{$u_c/c$} &
\colhead{$u_b/c$} &
\colhead{$\mathcal{A}_{\rm eff}$} &
\colhead{$v_A/c$} &
\colhead{$m_p/m_e$} &
\colhead{$\gamma_m/\Omega_e$} &
\colhead{$\theta_m$} &
\colhead{$ck_m/\omega_{pe}$} 
}
\startdata
Lu0.22&          50 & 0.2 & 0.044 & -0.176 & 1.46 & $6\times10^{-4}$   & 1836 & 0.17 & $79^{\circ}$ & 0.31\\
Lu0.26&          50 & 0.2 & 0.052 & -0.208 & 1.65 & $6\times10^{-4}$   & 1836 & 0.21 & $80^{\circ}$ & 0.34\\
Lu0.3 &          50 & 0.2 & 0.06  & -0.24 & 1.86  & $6\times10^{-4}$   & 1836 & 0.24 & $81^{\circ}$ & 0.38\\
Lu0.3$\beta50$ & 25 & 0.2 & 0.06  & -0.24 & 1.86  & $8.5\times10^{-4}$ & 1836 & 0.21 & $79^{\circ}$ & 0.41\\
Lu0.3m100      & 50 & 0.2 & 0.06  & -0.24 & 1.86  & $2.6\times10^{-3}$ & 100  & 0.24 & $81^{\circ}$ & 0.38\\
\enddata
\tablenotetext{{\rm a}}{See Section \ref{sec:s2.2} for the model naming convention.}
\end{deluxetable*}

\begin{figure*}[t]
\vskip 0.2cm
\hskip 0cm
\centerline{\includegraphics[width=1\textwidth]{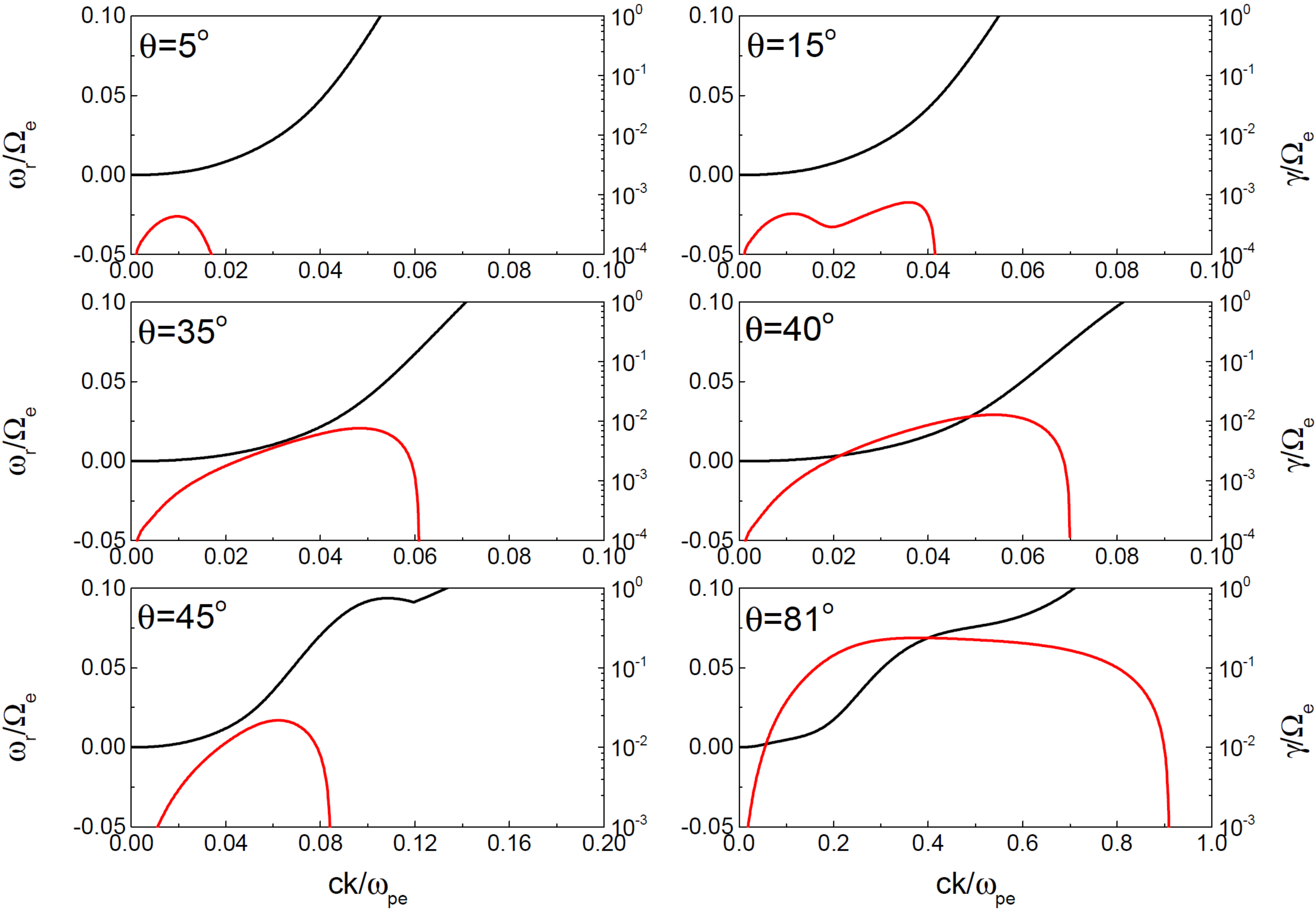}}
\vskip -0.2cm
\caption{Real frequency, $\omega_r$ (black), and growth rate, $\gamma$ (red), of the EBFI for the Lu0.3 model in Table \ref{tab:t1}, as a function of wavenumber, $k$, for different propagation angle, $\theta$. Note that the ranges of abscissa and ordinate differ in different panels.
\label{fig:f3}}
\end{figure*}

\begin{figure*}[t]
\vskip 0.2cm
\hskip 0cm
\centerline{\includegraphics[width=1\textwidth]{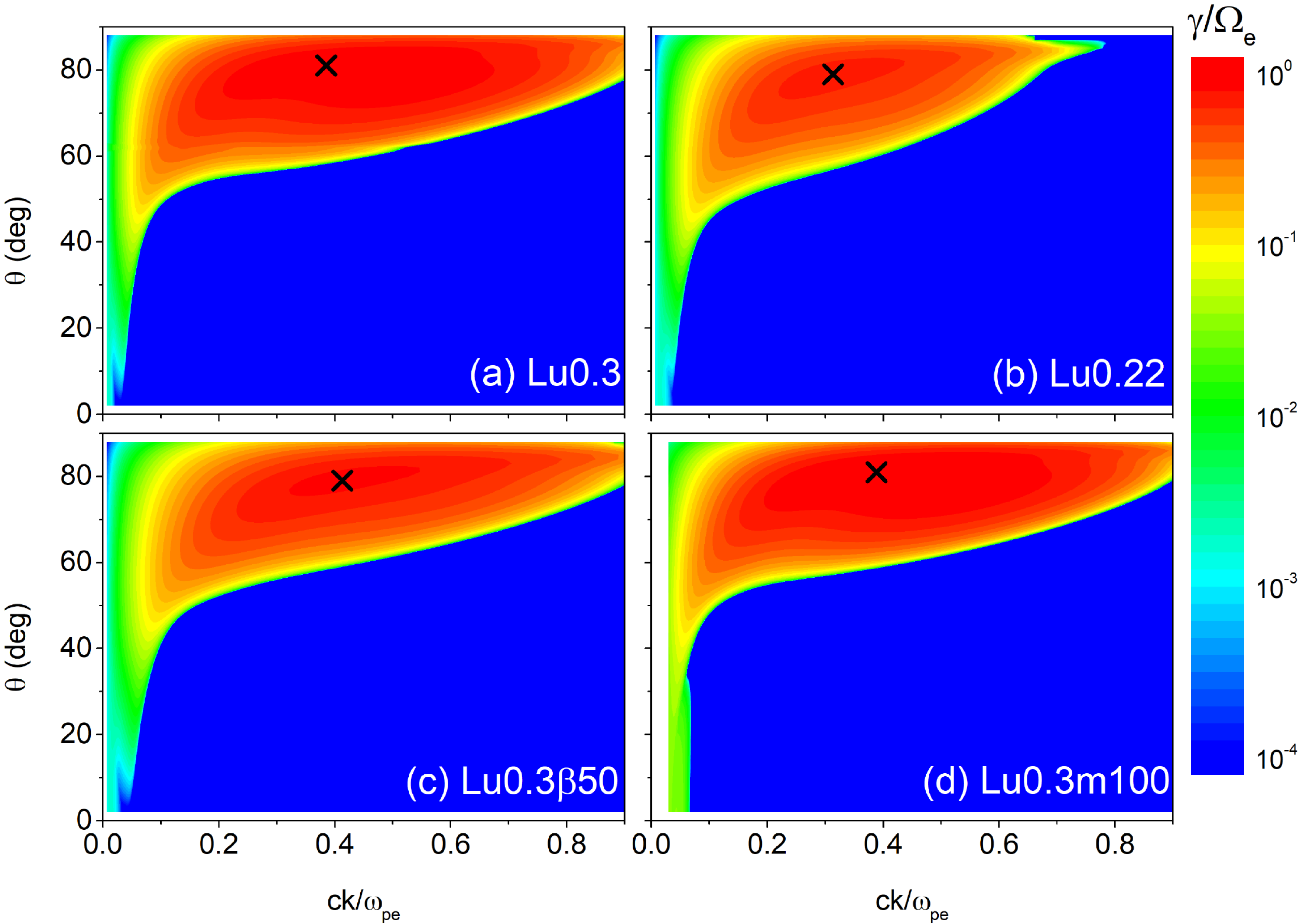}}
\vskip -0.2cm
\caption{Angular and wavenumber dependence of the growth rate, $\gamma$, of the EBFI for four models, (a) Lu0.3, (b) Lu0.22, (c) Lu0.3$\beta$50, and (d) Lu0.3m100, in Table \ref{tab:t1}. The X marks the location of the maximum growth rate in the $k$-$\theta$ plane.
\label{fig:f4}}
\end{figure*}

\subsection{Electron Beam Firehose Instability (EBFI)}\label{sec:s2.2}

In this subsection, we examine the EBFI in the ICM environment, induced by a drifting beam of electrons. Three populations of core electrons, beam electrons, and ions are involved, and we assume that all follow isotropic Maxwellian VDFs. We again restrict the analysis to the case of $\beta_e=\beta_p$, or equivalently $T_e=T_p$. Then, with the charge neutarlity and zero net current conditions, the analysis is reduced to a problem of six parameters, for instance, $T_e$, $n_c$, $n_b$, $u_{\rm rel}\equiv u_c-b_b$, $B_0$, and $m_p/m_e$. We specify the problem with five dimensionless quantities, $\beta$, $n_b/n_0$, $u_{\rm rel}/c$, $v_A/c$, and $m_p/m_e$, again using $\omega_{pe}$ to normalize $k$.

Emulating backstreaming electrons in the foot of a simulated shock in the ICM environment, specifically, the M3.0 model shock ($M_{\rm s}=3.0$, $\beta=100$) of KRH19, we adopt the model of $\beta=100$, $n_b/n_0=0.2$ (then, $n_c/n_0=0.8$), $u_{\rm rel}/c=0.3$ ($u_c=0.06$ and $u_b=-0.24$), $v_A/c=6\times10^{-4}$, and $m_p/m_e=1836$ as the fiducial model. We also consider four additional models to explore the dependence of the EBFI on $u_{\rm rel}/c$, $\beta$, and $m_p/m_e$, as listed in Table \ref{tab:t1}. The model name in the table has the following meaning. The first character `L' stands for `linear analysis'. The letter `u' is followed by $u_{\rm rel}/c$; the Lu0.3 model in the third row is the fiducial case. The models in the last two rows are appended by a character for the specific parameter and its value that is different from the fiducial value; the Lu0.3$\beta50$ model has $\beta=50$, and the Lu0.3m100 model has $m_p/m_e=100$. The last three columns of the table show $\gamma_m$, $\theta_m$ and $k_m$ of the fastest-growing mode.

To compare the characteristics of the EBFI with those of the ETAFI, we define an ``effective'' temperature anisotropy as follows. The ``effective'' parallel and perpendicular temperatures of the total (core plus beam) electron population are estimated as 
\begin{eqnarray}
\label{eq:e06}
\nonumber
T_{e \parallel}^{\rm eff} &=& \frac{m_e}{k_B n_0} \int d^3v \left(v_{\parallel}-\langle v_{\parallel}\rangle\right)^2 f_e
\nonumber \\
&=& T_e + \frac{m_e}{k_B}\left(u_c^2 \frac{n_c}{n_0}+u_b^2 \frac{n_b}{n_0}\right),
\nonumber \\
T_{e \perp}^{\rm eff} &=& \frac{m_e}{k_B n_0} \int d^3v \frac{v_{\perp}^2}{2} f_e 
= T_e,
\end{eqnarray}
where $f_e=f_c+f_b$. Note that
\begin{equation}
\label{eq:e07}
\langle v_{e \parallel}\rangle = \frac{1}{n_0} \int d^3v\ v_{\parallel} f_e=\frac{n_c}{n_0}u_c+\frac{n_b}{n_0}u_b=0,
\end{equation} 
with the zero net current condition in the ion rest frame. Then, the effective temperature anisotropy, arsing from the drift of electrons, is given as $\mathcal{A}_{\rm eff}=T_{e \parallel}^{\rm eff}/T_{e \perp}^{\rm eff}$; it is listed in the sixth column of Table \ref{tab:t1}.

\begin{deluxetable*}{ccccccccccc}[t]
\tablecaption{Model Parameters for the PIC Simulations of the EBFI\label{tab:t2}}
\tabletypesize{\small}
\tablecolumns{11}
\tablenum{2}
\tablewidth{0pt}
\tablehead{
\colhead{Model Name $^{\rm a}$} &
\colhead{$\beta_e$=$\beta_p$} &
\colhead{$n_b/n_0$} &
\colhead{$u_{c}/c$} &
\colhead{$u_{b}/c$} &
\colhead{$\mathcal{A}_{\rm eff}$} &
\colhead{$T_e=T_p [\rm K(keV)]$} &
\colhead{$m_p/m_e$} &
\colhead{$L_x=L_y [c/\omega_{pe}]$} &
\colhead{$\Delta x=\Delta y[c/\omega_{pe}]$} &
\colhead{$t_{\rm end} [\Omega_{\rm e}^{-1}]$}
}
\startdata
Su0.22&          50 & 0.2 & 0.044 & -0.176 & 1.46 & $10^8(8.6)$ & 100 & $100$ & 0.1 & $1000$ \\
Su0.26&          50 & 0.2 & 0.052 & -0.208 & 1.65 & $10^8(8.6)$ & 100 & $100$ & 0.1 & $1000$ \\
Su0.3&           50 & 0.2 & 0.06  & -0.24  & 1.86 & $10^8(8.6)$ & 100 & $100$ & 0.1 & $1000$ \\
Su0.3$\beta50$ & 25 & 0.2 & 0.06  & -0.24  & 1.86 & $10^8(8.6)$ & 100 & $100$ & 0.1 & $1000$ \\
Su0.3m400&       50 & 0.2 & 0.06  & -0.24  & 1.86 & $10^8(8.6)$ & 400 & $100$ & 0.1 & $1000$ \\
\enddata
\tablenotetext{{\rm a}}{See Section \ref{sec:s3.1} for the model naming convention.}
\end{deluxetable*}

\begin{figure*}[t]
\vskip 0.2cm
\hskip 0cm
\centerline{\includegraphics[width=1\textwidth]{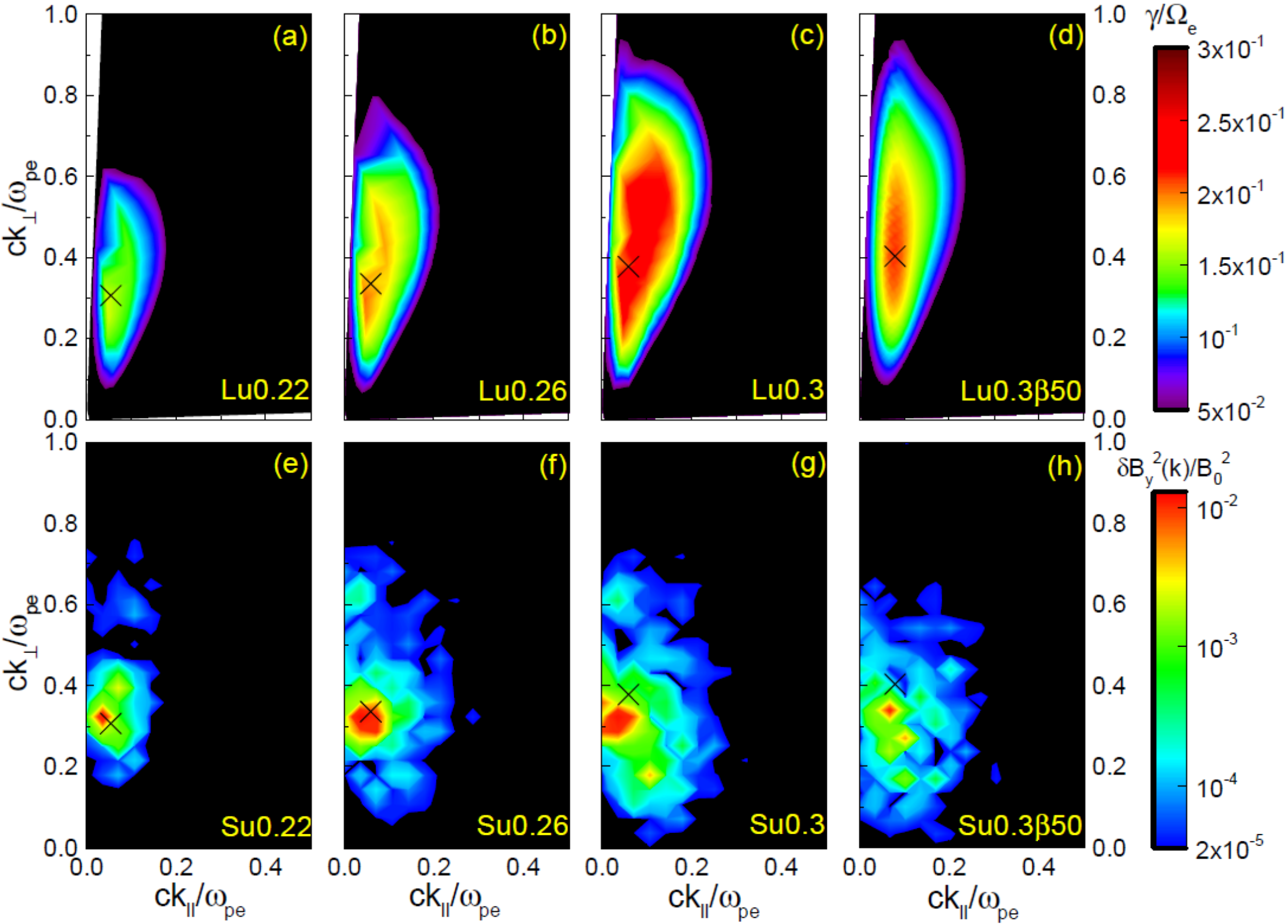}}
\vskip -0.2cm
\caption{Top panels: the growth rate of the EBFI, $\gamma(\mathbf{k})/\Omega_e$, in the $k_\parallel$-$k_\perp$ plane for four models in Table \ref{tab:t1}, calculated by the linear analysis in Section \ref{sec:s2.2}. Bottom panels: the square of the Fourier transformation of the $y$-magnetic field fluctuations, $\delta B_y^2(\mathbf{k})/B_0^2$, in the $k_\parallel$-$k_\perp$ plane for four models in Table \ref{tab:t2}, estimated at $\Omega_{e}t = 5$, from PIC simulations. Note that the color bar of $\gamma(\mathbf{k})/\Omega_e$ is in the linear scale, while that of $\delta B_y^2(\mathbf{k})/B_0^2$ is in the logarithmic scale. The parameters of Lu models are identical to those of their respective Su models, except that $m_p/m_e=1836$ for Lu models while $m_p/m_e=100$ for Su models. The X marks the location of the maximum linear growth rate, $\gamma_m$, of the Lu models.
\label{fig:f5}}
\end{figure*}

Figure \ref{fig:f3} shows the normalized real frequency, $\omega_r/\Omega_e$ (black line), and the normalized growth rate, $\gamma/\Omega_e$ (red line), for the Lu0.3 model of the EBFI, as a function of $ck/\omega_{pe}$ for different $\theta$. This model has $\mathcal{A}_{\rm eff}=1.86$, which is the same as $\mathcal{A}$ of the ETAFI model of Figure \ref{fig:f1}. The magnitude of $\gamma$ and the unstable wavenumber range in Figure \ref{fig:f3} are comparable to those in Figure \ref{fig:f1}, and in both the figures, $\gamma$ increases with increasing $\theta$. For the Lu0.3 model, the maximum growth, $\gamma_m/\Omega_e=0.24$, appears at $\theta_m\approx81^{\circ}$, and at this angle, modes with a broad range of $ck/\omega_{pe}\sim0.2-0.8$ have $\gamma$ close to $\gamma_m$. 
The wavenumber of $\gamma_m$ is $ck_m/\omega_{pe}\approx0.38$ for the Lu0.3 model, smaller than $ck_m/\omega_{pe}\approx0.49$ at $\theta_m\approx85^{\circ}$ in Figure \ref{fig:f1}.
The more notable difference is that fast-growing oblique modes of the EBFI have $\omega_r \neq 0$, while those of the ETAFI have $\omega_r =0$. However, $\omega_r < \gamma$, for most of the modes; for the Lu0.3 model, $\gamma_m/\Omega_e\approx0.24$ and $\omega_r/\Omega_e\approx0.06$ 
at $k_m$ and $\theta_m$, and hence, $(\omega_r/k_m)/\gamma_m\approx0.66\ c/\omega_{pe} \ll \lambda_m(\equiv 2\pi/k_m)\approx16.5\ c/\omega_{pe}$, that is, the fastest-growing mode propagates the distance much smaller than its wavelength during the linear grow time of $1/\gamma_m$. It means that EM fluctuations grow much faster than they propagate. Thus, the oblique mode of the EBFI may be regarded as ``nearly phase-standing", while the oblique mode of the ETAFI is truly nonpropagating.

Figure \ref{fig:f4} demonstrates the effects of $u_{\rm rel}/c$, $\beta$, and $m_p/m_e$ on the growth rate, $\gamma$, of the EBFI in the $k$-$\theta$ plane. The black ``X'' denotes the location $(k_m,\theta_m)$ of the fastest-growing mode. The comparison of the Lu0.22 and Lu0.3 (also Lu0.26, although not shown) models indicates that for larger $u_{\rm rel}$ (i.e., larger $\mathcal{A}_{\rm eff}$), $\gamma$ peaks at larger $k_m$ and larger $\theta_m$. This is consistent with the result of the ETAFI, shown in the right panels of Figure \ref{fig:f2}. The panels (a) and (c), which compare the Lu0.3 and Lu0.3$\beta$50 models, illustrate that for smaller $\beta$, $k_m$ is larger, while $\theta_m$ is smaller. Such $\beta$ dependence is also seen in the case of the ETAFI, shown in the left panels of Figure \ref{fig:f2}. The panels (a) and (d), which compare the Lu0.3 and Lu0.3m100 models, manifest that $m_p/m_e$ is not important, especially at high oblique angles of $\theta\gtrsim40^{\circ}$, as in the ETAFI. In summary, these characteristics of the EBFI are similar to those of the ETAFI. Hence, we expect that the EBFI would behave similarly to the ETAFI.

\section{PIC Simulations of EBFI}\label{sec:s3}

\subsection{Simulation Setup}\label{sec:s3.1}

\begin{figure*}[t]
\vskip 0.2cm
\hskip 0cm
\centerline{\includegraphics[width=1.1\textwidth]{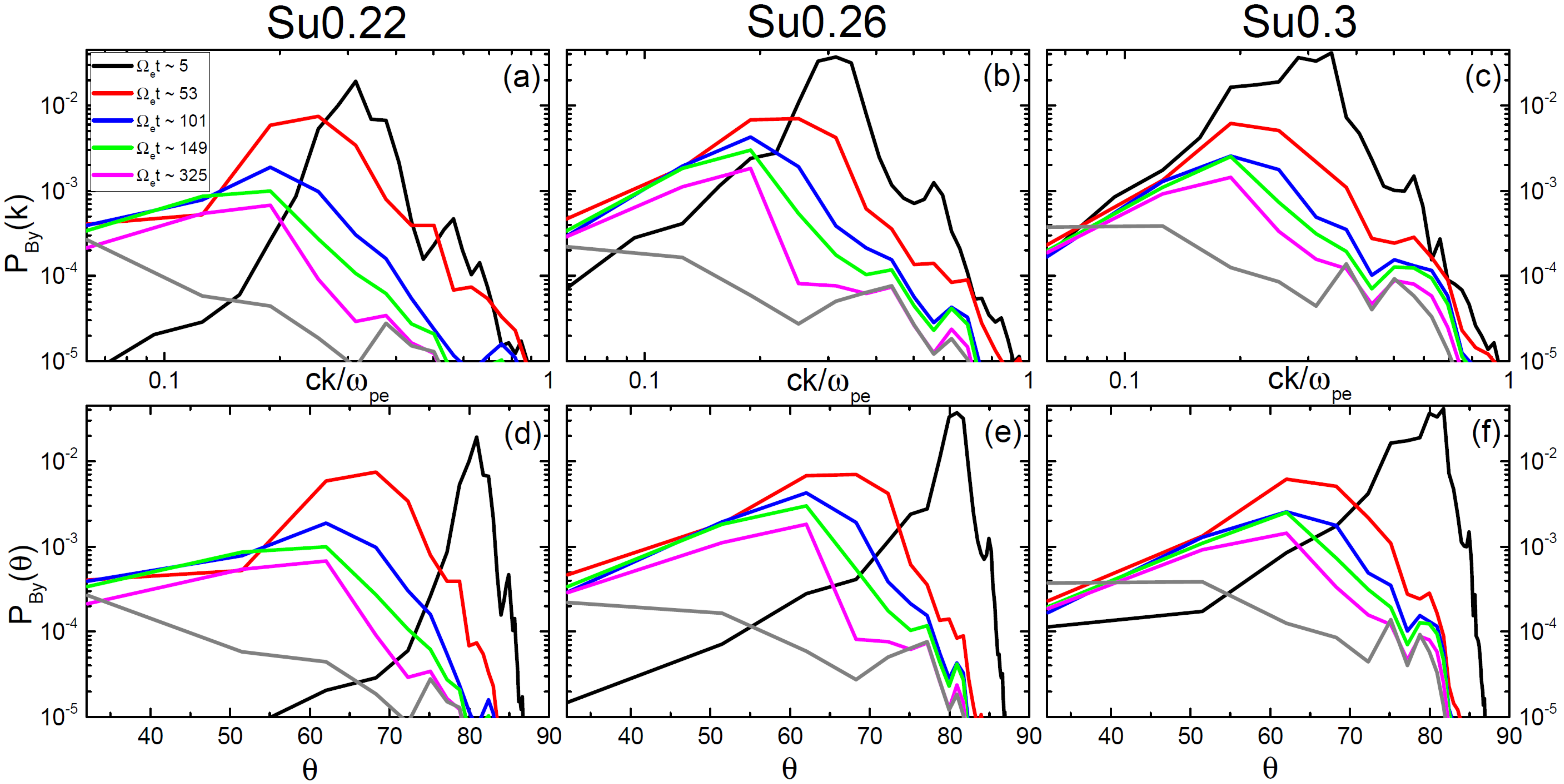}}
\vskip -0.2cm
\caption{Time evolution of $P_{B_y}(k)$ (top panels) and $P_{B_y}(\theta)$ (bottom panels), the power of $\delta B_y/B_0$, as a function of $k$ and $\theta$, at different times in the PIC simulations for the Su0.22, Su0.26 and Su0.3 models in Table \ref{tab:t2}. The gray lines show the power at a later time, $\Omega_e t \sim 500$.
\label{fig:f6}}
\end{figure*}

To further explore the development and evolution of the EBFI in the foot of weak $Q_{\perp}$-shocks in the ICM, we study the instability through 2D PIC simulations. We consider the setup equivalent to that of Section \ref{sec:s2.2}; electrons, described with an isotropic Maxwellian VDF, drift along the direction of the background magnetic field, $\mathbf B_0 = B_0 \hat{z}$. In fact, \citet{guo2014b} performed similar PIC simulations to describe the triggering instability and the properties of excited upstream waves, seen in their shock study. The difference is that in their simulations, the beam electrons are drifting within the maximum pitch angle and have a power-law energy distribution.

The PIC simulations were performed using TRISTAN-MP, a parallelized EM PIC code \citep{buneman1993, spitkovsky2005}. All the three components of the particle velocity and the EM fields are calculated within a periodic box. As in Section \ref{sec:s2.2}, the background plasma consists of core electrons, beam electrons, and ions. The core and beam electron populations drift, satisfying the zero net current condition, while the ion population is at rest. The simulation domain is in the $z-x$ plane. Again, the case of $\beta_e=\beta_p$, or equivalently $T_e=T_p$, is considered.

Parallel to the models for the linear analysis considered in Section \ref{sec:s2.2}, we ran simulations for the five models listed in Table \ref{tab:t2}. The model name in the first column has the same meaning as that in Table \ref{tab:t1}, except that the first character `S' stands for `simulation'. Su0.3 in the third row is the fiducial model; $\beta=100$, $n_b/n_0=0.2$, $u_{\rm rel}/c=0.3$, $T_e=T_p=8.6$ keV, and $m_p/m_e=100$. 
Again, this model is to intended to reproduce the upstream condition of the M3.0 model shock of KRH19. Note that here $m_p/m_e=100$ is used to speed up the simulations, but the early, linear-stage evolution of fast-growing oblique modes should be insensitive to the mass ratio, as mentioned above. Four additional models are considered to explore the dependence on $u_{\rm rel}/c$, $\beta$, and $m_p/m_e$.

The simulation domain is represented by a square grid of size $L_z=L_x=100\ c/\omega_{pe}$, which consists of cells of $\Delta z =\Delta x = 0.1\ c/\omega_{pe}$. In each cell, 200 particles (100 for electrons and 100 for ions) are placed. The time step is $\Delta t = 0.045\ [\omega_{pe}^{-1}]$, and the simulations ran up to $t_{\rm end} = 1000\ \Omega_{\rm e}^{-1}$.

\subsection{Simulations Results}\label{sec:s3.2}

\begin{figure*}[t]
\vskip 0.2cm
\hskip 0cm
\centerline{\includegraphics[width=1.1\textwidth]{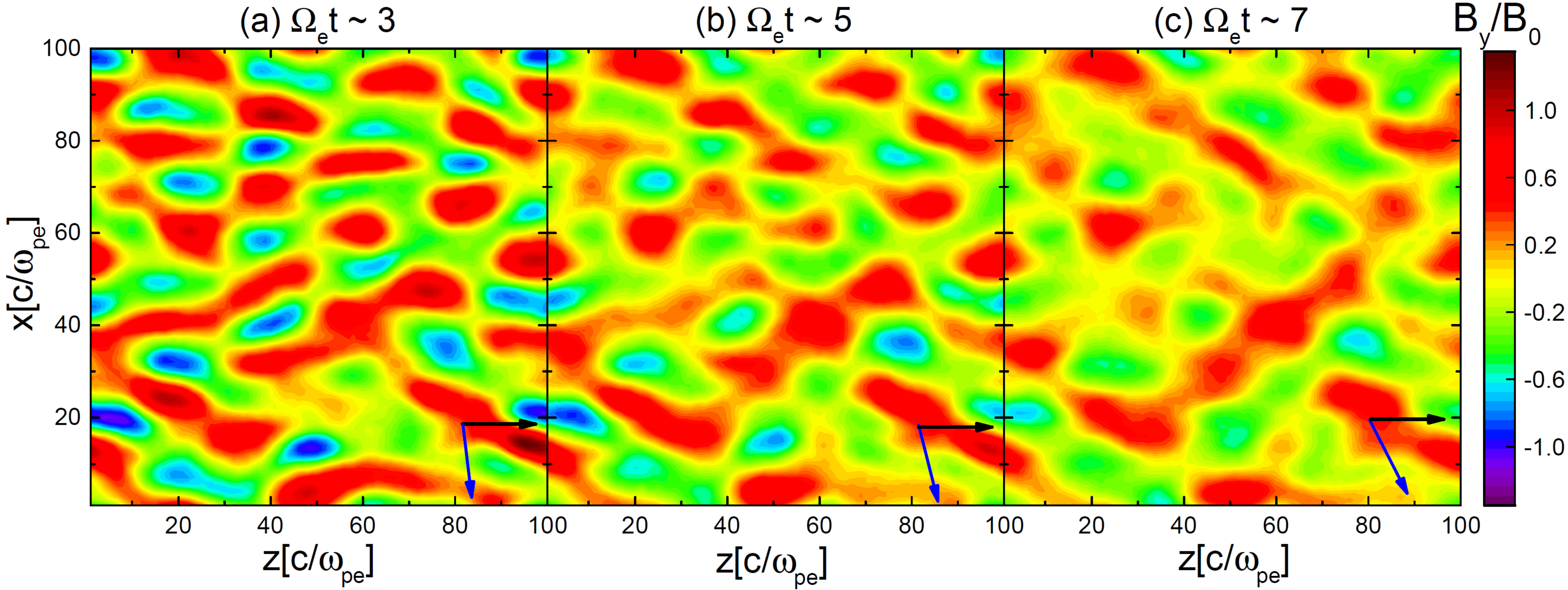}}
\vskip -0.2cm
\caption{Fluctuations of the $y$-magnetic field, $\delta B_y/B_0$, at three different times in the PIC simulation for the Su0.3 model. The black arrows draw the background magnetic field direction, while the blue arrows point the wavevector directions of the peaks of $P_{B_y}(\mathbf{k})$.
\label{fig:f7}}
\end{figure*}

\begin{figure}[t]
\vskip 0.2cm
\hskip -0.2cm
\centerline{\includegraphics[width=0.55\textwidth]{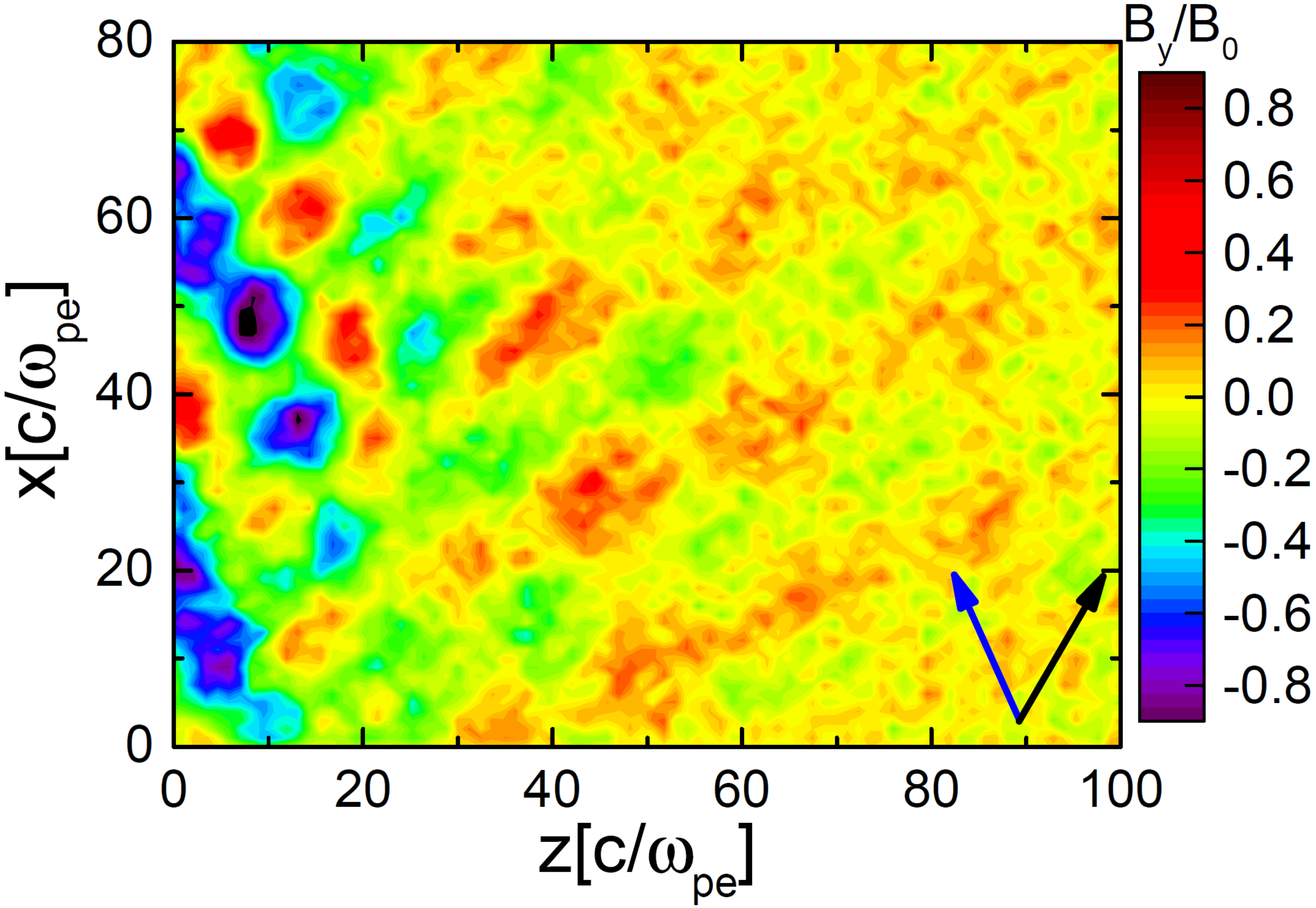}}
\vskip -0.2cm
\caption{Fluctuations of the $y$-magnetic field, $\delta B_y/B_0$, at $\Omega_e t\sim5000$ in the PIC simulation of the M3.0 model shock of KRH19. The black arrow draw the background magnetic field direction, while the blue arrow points the wavevector direction of the peak of $P_{B_y}(\mathbf{k})$.
\label{fig:f8}}
\end{figure}

As in the ETAFI (see the introduction), for fast-growing oblique modes of the EBFI, the magnetic field fluctuations are induced predominantly along the direction perpendicular to both $\mathbf{k}$ and $\mathbf{B_0}$, i.e., along the $y$ axis in our geometry.\footnote{We confirmed it in the simulations, although we do not explicitly show it here.} 
Hence, below, we present the simulation results associated with $\delta B_y$ to describe the evolution of the EBFI. With its Fourier transformation, $\delta B_y(\mathbf{k})$, we first compare $\ln(\delta B_y^2(\mathbf{k})/B_0^2)$, calculated in the PIC simulations, with the linear growth rate, $\gamma(\mathbf{k})$, described in Section \ref{sec:s2.2}, since $\delta B_y(\mathbf{k}) \propto \exp(\gamma(\mathbf{k}))$ in the linear regime. 
Figure \ref{fig:f5} shows such comparison between $\gamma(\mathbf{k})$ of the linear analysis models in Table \ref{tab:t1} (top panels) and $\delta B_y^2(\mathbf{k})/B_0^2$ of their respective simulation models in Table \ref{tab:t2} (bottom panels).
Here, $\delta B_y(\mathbf{k})$ is at $\Omega_e t= 5$, close to the linear growth time of the fastest-growing mode, $\Omega_e/\gamma_m$. In the linear analysis, the fastest-growing mode occurs at $k_mc/\omega_{pe} \sim 0.31-0.41$ and $\theta_m \sim 79^{\circ}-81^{\circ}$ (see Table \ref{tab:t2}), which corresponds to the positions of the black ``X'' marks in the figure. The figure demonstrates a fair consistency between the simulations and the linear analysis. The bottom panels show that $\ln(\delta B_y^2(\mathbf{k})/B_0^2)$ is substantial in the portion of the $k_\parallel$-$k_\perp$ plane where the growth rate is substantial. 
In the Su0.22 and Su0.26 models, the peak of $\delta B_y^2(\mathbf{k})/B_0^2$ agrees reasonably well with the location of the X mark.
In the Su0.3 and and Su0.3$\beta$50 models, on the other hand, the peak shifts a little to the lower left direction of the X mark, possibly a consequence of the nonlinear evolution of the instability (see below).

Although the Su0.3m400 model is not presented in Figure \ref{fig:f5}, we find that the distribution of $\ln(\delta B_y^2(\mathbf{k})/B_0^2)$ in the $k_\parallel$-$k_\perp$ plane coincides well with that of the Su0.3 model. This confirms that the development of the EBFI is not sensitive to $m_{p}/m_{e}$ in the nonlinear regime as well as in the linear regime.

As described in the introduction, previous studies of the ETAFI have shown that as the instability develops, the magnetic field fluctuations inversely cascade toward longer wavelengths and smaller $\theta$, and that the scattering of electrons by excited waves reduces the temperature anisotropy and the ETAFI-induced waves decay \citep[e.g.,][]{camporeale2008,hellinger2014}. We expect a similar inverse cascade for the EBFI-driven magnetic field fluctuations as well.
In addition, excited waves will disperse the electron beam, resulting in the decrease of the relative drift speed and eventually leading to the damping of the magnetic field fluctuations with time.

To describe the evolution of the EBFI, we examine the magnetic power spectra, $P_{B_y}(k)$ and $P_{B_y}(\theta)$, defined with the following relations,
\begin{equation}
\frac{\delta B_y^2}{B_0^2} = \int P_{B_y}(k)d\ln k = \int P_{B_y}(\theta)d\theta.
\end{equation}
Note that $P_{B_y}(k)=(\delta B_y^2(k)/B_0^2)k^2$. Figure \ref{fig:f6} shows the time evolution of $P_{B_y}(k)$ and $P_{B_y}(\theta)$ for three Su models. We first see that at the early time of $\Omega_e t= 5$, the peaks of $P_{B_y}(k)$ and $P_{B_y}(\theta)$ occur at the values close to those predicted in the linear analysis, $k_m$ and $\theta_m$ (see the discussion above). The figure also demonstrates that the magnetic power transfers to smaller $k$ and smaller $\theta$; such inverse cascade continues to $kc/\omega_{pe} \sim 0.2$ (corresponding wavelength is $\lambda \sim 30 c/\omega_{pe}$) and $\theta \sim 60^{\circ}$ at $\Omega_{\rm e}t \sim 300$. Eventually, the magnetic power decays away in the timescale of $\Omega_{\rm e}t \sim 500$, indicating that the modes of long wavelengths with $\lambda \gg \lambda_m$ are not produced by the EBFI.

A similar evolutionary behavior of the magnetic field fluctuations, that is, the inverse cascade followed by the decay, was observed in the simulations of weak $Q_\perp$-shocks in the high-$\beta$ ICM plasmas presented by KRH19. In the shocks, however, the beam of SDA-reflected electrons is, although fluctuating, continuously supplied from the shock ramp, persistently inducing the instability. As a consequence, the magnetic field fluctuations exhibit an oscillatory behavior, showing the rise of the instability, followed by the inverse cascade of the magnetic power, and then the decay of turbulence (see  Figure 9 of KRH19). The period of such oscillations is $\Omega_{\rm e}t \sim 500-1000$, close to the decay time scale of the EBFI. Even in the shocks with a continuous stream of reflected electrons, the modes of long wavelengths ($\lambda \gg \lambda_m$) do not develop, as shown in KRH19. 

The linear analysis of the EBFI in Section 2.2 indicates that fast-growing oblique modes, although they are propagating with $\omega_r \ne 0$, have mostly $\omega_r < \gamma$. So these modes are ``effectively" phase-standing, similar to the oblique nonpropgating modes excited by the ETAFI. Figure \ref{fig:f7} shows the spatial distribution of $\delta B_y/B_0$ at three different times covering almost one linear growth time in the PIC simulation for the Su0.3 model. The figure demonstrates that the oblique modes induced by the EBFI are indeed almost nonpropgating. 
It also illustrates visually that the peak of $P_{B_y}(k)$ shifts gradually toward longer wavelength and smaller $\theta$, while the magnetic field fluctuations decay.

Figure \ref{fig:f8} shows the spatial distribution of $\delta B_y/B_0$ in the foot of a shock, which is taken from the PIC simulation for the M3.0 model shock ($M_{\rm s}=3.0$, $\theta_{\rm Bn}=63^{\circ}$) reported by KRH19. The strong waves in the shock ramp at the left-hand side of the figure are whistlers excited by reflected ions; obviously they are absent in our periodic-box simulations for the EBFI. The oblique waves in the region, $x/[c/\omega_e]>30$, on the other hand, are well compared with those in Figure \ref{fig:f7}. 
In particular, the wavelength and $\theta$ of the peak of $P_{B_y}$ are comparable to those in Figure \ref{fig:f7}(c). By considering the origin of the instability and also the similarity between Figures \ref{fig:f7} and \ref{fig:f8}, we conduce that it should be the EBFI due to the beam of SDA-reflected electrons that operates in the foot of $Q_\perp$-shocks in the ICM. We also argue that the upstream waves excited by the EBFI, although they have non-zero $\omega_r$, can be regarded as almost phase-standing.

\section{Summary}\label{sec:s4}

Recent studies for the electron preacceleration in weak $Q_\perp$-shocks in the high-$\beta$ ICM plasmas suggested that the temperature anisotropy of $T_{\parallel}>T_{\perp}$ due to SDA-reflected electrons generates oblique waves in the shock foot via the EFI, i.e., the ETAFI \citep[][KRH19]{guo2014a, guo2014b}. The electrons can be effectively trapped between the shock ramp and these upstream waves, and hence continue to gain energy through multiple cycles of SDA. Those studies compared the properties of the excited upstream waves in shock simulations with the results of the linear analysis and PIC simulations of the ETAFI driven by anisotropic bi-Maxwellian electrons \citep[e.g.][]{gary2003,camporeale2008}. In the $Q_\perp$ ICM shocks, however, the instability due to SDA-reflected electrons is expected to be more like the electron heat flux instability driven by a drifting beam, i.e., the EBFI, since the electrons stream along the background magnetic field with small pitch angles, and hence they would behave similar to the electrons of a drifting beam rather than bi-Maxwellian electrons.

To describe the nature of the upstream waves excited in the shock foot, we here studied the EFI in two different forms: (1) the ETAFI induced by the electrons of a bi-Maxwellian VDF with the temperature anisotropy, $\mathcal{A}=T_{e\parallel}/T_{e\perp}>1$, and (2) the EBFI induced by the electrons of a drifting beam with an isotropic Maxwellian VDF and the relative drift speed, $u_{\rm rel}$. We carried out the kinetic linear analysis of both types of the EFI in Section \ref{sec:s2} and the 2D PIC simulations of the EBFI in Section \ref{sec:s3}.

The main findings can be summarized as follows:

1. In the EBFI, an effective temperature anisotropy, $\mathcal{A}_{\rm eff}$, can be defined (see Section \ref{sec:s2.2}); $\mathcal{A}_{\rm eff}$ is larger for larger $u_{\rm rel}$. In the linear analysis, the characteristics of the EBFI are similar to those of the ETAFI, if $\mathcal{A}_{\rm eff}$ is similar to $\mathcal{A}$.

2. For both the EFI instabilities, the oblique modes with a large propagation angle $\theta$ grow faster, having a higher growth rate $\gamma$, than the parallel modes with a small $\theta$. 
For the Lu0.3 model, the fiducial model of the EBFI, for example, $ck_m/\omega_{pe}=0.38$ and $\theta_m=81^\circ$.
 
3. The growth rates of both the instabilities increase with the increasing plasma beta $\beta$ \citep[see, e.g.,][]{gary2003}.
For higher $\beta$, the fastest-growing mode occurs at smaller $k_m$ and larger $\theta_m$.

4. Naturally, the growth rate increases with increasing $\mathcal{A}$ or $u_{\rm rel}$. 
For larger $\mathcal{A}$ or $u_{\rm rel}$, the fastest-growing mode occurs at larger $k_m$ and larger $\theta_m$. 

5. In both the instabilities, the growth rate of fast-growing oblique modes is insensitive to $m_p/m_e$, for a sufficiently large mass ratio (i.e., $m_p/m_e\gtrsim100$), as shown in previous studies \citep[e.g.,][]{gary2003}.

6. The fast-growing oblique modes excited by the ETAFI are nonpropagating with $\omega_r =0$ (zero real frequency), while those excited by the EBFI have $\omega_r\ne0$, but $\omega_r<\gamma$. Hence, the fast-growing modes of the EBFI is nearly phase-standing, even though they are propagating.

7. The PIC simulations of the EBFI presented in Section \ref{sec:s3} show that the time evolution of the magnetic field fluctuations induced by this instability is consistent with the prediction of the linear analysis given in Section \ref{sec:s2.2} and also with the results for the ETAFI reported by \citet{camporeale2008} and \citet{hellinger2014}. The oblique, almost nonpropagating modes inverse-cascade in time to the modes with smaller wavenumbers, $k$, and smaller propagation angles, $\theta$. The scattering of electrons by these waves reduces the beam strength, which in turn leads to the damping of the waves. As a result, the modes of long wavelengths with $\lambda \gg \lambda_m \sim 15-20 c/\omega_{pe}$ are not produced by the EBFI.

As argued by KRH19, without longer waves that can scatter higher energy electrons, the Fermi I-like preacceleration in the shock foot may not proceed to all the way to $p_{\rm inj}$ at weak $Q_\perp$-shocks in the ICM. However, \citet{trotta2019} and \citet{kobzar2019} have recently shown that, if the simulation volume is large enough to include ion-scale perturbations, the shock surface rippling caused by the Alfv\'en ion cyclotron instability can generate multiscale waves, leading the electron injection to DSA. Additional elements, such as pre-existing fossil CR electrons and/or pre-exiting turbulence on kinetic plasma scales in the ICM, may also facilitate the electron injection to DSA.

\acknowledgments
S.K., J.-H. H., \& D.R. were supported by the National Research Foundation of Korea (NRF) through grants 2016R1A5A1013277 and 2017R1A2A1A05071429. J.-H. H. was also supported by the Global PhD Fellowship of the NRF through grant 2017H1A2A1042370. H.K. was supported by the Basic Science Research Program of the NRF through grant 2017R1D1A1A09000567. This research was supported in part by the National Science Foundation under Grant No. NSF PHY-1748958.

\appendix

\section{Linear Dispersion Relation}

\restartappendixnumbering{}

For the general VDF of constituent species, $f_a$, the $ij$-component of the dielectric tensor, $\epsilon_{ij}$, can be written as
\begin{eqnarray}
\label{eq:eA1}
\nonumber
&& \epsilon_{ij} = \delta_{ij}+\sum_{a } \frac{\omega_{pa}^2}{\omega^2} \int d^3v 
\bigg[ v_{\parallel} \bigg(\frac{\partial}{\partial v_{\parallel}}-\frac{v_{\parallel}}{v_{\perp}} \frac{\partial}{\partial v_{\perp}} \bigg) 
f_a \hat{b}_i \hat{b}_j \\
\nonumber
&&+\sum_{n=-\infty}^{n=\infty} \frac{V_i V_j^*}{\omega-n\Omega_a-k_{\parallel}v_{\parallel}}
\bigg(\frac{\omega-k_{\parallel}v_{\parallel}}{v_{\perp}} \frac{\partial}{\partial v_{\perp}}+k_{\parallel}\frac{\partial}{\partial v_{\parallel}} \bigg) f_a \bigg],\\
\end{eqnarray}
where
\begin{eqnarray}
\label{eq:eA2}
\nonumber
V_i &=& \bigg(v_{\perp}\frac{nJ_n(b_{\perp})}{b_{\perp}}, -iv_{\perp}J_n^{\prime}(b_{\perp}), v_{\parallel} J_n(b_{\perp})\bigg), \\
b_{\perp} &=& \frac{k_{\perp}v_{\perp}}{\Omega_a},
\quad \hat{b}_i=\frac{B_{0i}}{B_0},
\end{eqnarray}
and $J_n$ is the Bessel function. Here, $*$ denotes the complex conjugate.

For the drifting bi-Maxwellian VDF given in Equation (\ref{eq:e01}), it can be shown that
\begin{eqnarray}
\label{eq:eA3}
\nonumber
&& \bigg(\frac{\omega-k_{\parallel}v_{\parallel}}{v_{\perp}} \frac{\partial}{\partial v_{\perp}}+k_{\parallel}\frac{\partial}{\partial v_{\parallel}} \bigg) f_a \\
\nonumber
&&=-\frac{2}{\pi^{3/2}\alpha_{a \perp}^4\alpha_{a \parallel}}\bigg[\omega-k_{\parallel}u_a+\bigg(\frac{T_{a\perp}}{T_{a\parallel}}-1\bigg)k_{\parallel}(v_{\parallel}-u_a)\bigg] \\
\nonumber
&& \quad \times
\exp\left[-\frac{v_{\perp}^2}{\alpha_{a \perp}^2}-\frac{(v_{\parallel}-u_a)^2}{\alpha_{a \parallel}^2}\right], \\
&& \int d^3v\, v_{\parallel}\bigg(\frac{\partial}{\partial v_{\parallel}}-\frac{v_{\parallel}}{v_{\perp}} \frac{\partial}{\partial v_{\perp}} \bigg) f_a =\frac{T_{a\perp}}{T_{a\parallel}}-1.
\end{eqnarray}
Then, using the plasma dispersion function, $Z(\zeta)$, and the related identities \citep{Fried1961},
\begin{eqnarray}
\label{eq:eA4}
\nonumber
Z(\zeta) &=& \int_{-\infty}^{\infty} \frac{dy}{\pi^{1/2}} \frac{e^{-y^2}}{y-\zeta}, \\
\nonumber
-\frac{Z^{\prime}(\zeta)}{2} &=& \int_{-\infty}^{\infty} \frac{dy}{\pi^{1/2}} \frac{ye^{-y^2}}{y-\zeta}, \\
\nonumber
-\frac{\zeta Z^{\prime}(\zeta)}{2} &=& \int_{-\infty}^{\infty} \frac{dy}{\pi^{1/2}} \frac{y^2 e^{-y^2}}{y-\zeta}, \\
\frac{1}{2}\left[1-\zeta^2 Z^{\prime}(\zeta)\right]
&=& \int_{-\infty}^{\infty} \frac{dy}{\pi^{1/2}} \frac{y^3 e^{-y^2}}{y-\zeta},
\end{eqnarray}
and the integrals involving the Bessel functions,
\begin{eqnarray}
\label{eq:eA5}
\nonumber
2 \int_{0}^{\infty} d\chi\, \chi\, e^{-\chi^2}\, J_n^2(b\chi) = && I_n(\lambda)\,e^{-\lambda}, \\
\nonumber
\\
\nonumber
2 \int_{0}^{\infty} d\chi\, \chi^3\, e^{-\chi^2}\, J_n^2(b\chi) = && \left[\lambda\,I_n(\lambda)\,e^{-\lambda}\right]^{\prime}, \\
\nonumber
2 \int_{0}^{\infty} d\chi\, \chi^2\, e^{-\chi^2}\, J_n(b\chi)\,J_n^{\prime}(b\chi) = 
&& \frac{b}{2} \left[I_n(\lambda)\,e^{-\lambda}\right]^{\prime}, \\
\nonumber
4 \int_{0}^{\infty} d\chi\, \chi^3\, e^{-\chi^2}\, [J_n^{\prime}(b\chi)]^2
= && \frac{n^2 I_n(\lambda)\,e^{-\lambda}}{\lambda} \\
-2\lambda\, && \left[I_n(\lambda)\,e^{-\lambda}\right]^{\prime}, 
\end{eqnarray}
where $\lambda=b^2/2$, $\epsilon_{ij}$ in Equations (\ref{eq:e03}) and (\ref{eq:e04}) can be derived.

\end{document}